\documentclass[12pt,a4paper]{article}
\usepackage{amsmath,amssymb}
\usepackage{graphicx,float,colordvi,url,wrapfig}
\usepackage{color} 

\def\Tr{{\rm Tr}} 
\def\Det{{\rm Det}} 
\def\Log{{\rm Log}}   
 
\def\p{\partial} 

\def\nn{\nonumber\\} 
\def\black{\textcolor{black}} 
\def\red{\textcolor{black}} 
\def\blue{\textcolor{black}}

\begin{document} 

\begin{titlepage}

\vspace*{5mm}

\begin{center}
{\large \bf Bose-Einstein condensation in the Rindler space}

\vspace*{8mm}

\normalsize
{\large Shingo Takeuchi}

\vspace*{6mm} 

\textit{The Institute for Fundamental Study ``The Tah Poe Academia Institute''}\\
\vspace*{0.10 cm}
\textit{Naresuan University Phitsanulok 65000, Thailand}

\vspace*{2.0 mm}

\end{center}

\vspace*{5mm}
\begin{abstract}
Based on the Unruh effect, 
we calculate the critical acceleration for the Bose-Einstein condensation 
in a free complex scalar field 
at finite density in the Rindler space. 
%---
Our model corresponds to an ideal gas 
performing constantly accelerating motion 
in a Minkowski spacetime at zero temperature, 
where the gas is composed of the complex scalar particles, 
and is supposed to be in a thermal bath at some temperature by the Unruh effect. 
%---
%Then, while the BEC is formed at low acceleration, 
%there is be no condensation at high acceleration 
%for the thermal effect by the Unruh effect. 
%---
The critical Unruh temperature we obtain agrees with the usual result 
in the 4D Euclid space at finite temperature.
\end{abstract}
\end{titlepage} 

\newpage

%%%%%%%%%%%%%%%%%%%%%%%%%%%%%%%%%%%%%%%%%%%%%%%%%%%%%%% 
\section{Introduction}
\label{vnrver}
%%%%%%%%%%%%%%%%%%%%%%%%%%%%%%%%%%%%%%%%%%%%%%%%%%%%%%%

Recently, the Bose-Einstein condensation (BEC) is getting a lot of attention 
as the quantum fluid 
for the test of the analogy between 
sound waves in quantum fluids and scalar field fluctuations 
in curved spacetimes~\cite{Unruh:1980cg}. 
Based on this analogy, 
some pseudo experiments for gravity and cosmology are also possible in the quantum fluids, 
and we could expect new progress and insights 
in the quantum theory and quantum behavior of the gravity. 
%---

Next, in order to understand the quantum phenomena 
in the gravity and cosmology 
such as the Hawking radiation~\cite{Hawking:1974sw} 
and particle creation~\cite{Novello:2002fc,Barcelo:2005fc}, 
the Unruh effect~\cite{Davies:1974th,Unruh:1976db,Unruh:1983ac}\cite{Crispino:2007eb} 
is very important, 
which is the prediction 
that one moving in the Minkowski spacetime 
with a constant acceleration experiences 
the spacetime as a thermal-bath 
with the Unruh temperature, 
$T_U = \hbar\, a/(2\pi c \,k_B) \approx 4 \times 10^{-23}\,a/({\rm cm}/{\rm s}^2)\,[{\rm K}]$, 
where $a$ is the acceleration. 
%---
Currently, various experimental attempts 
in the condensed matter 
to observe analogous gravitational phenomena   
are being invented~(see Ref.\cite{Kurita:2008fb} for example). 
In particular, there are some experimental attempts 
to detect the Unruh effect 
in the BEC~\cite{Retzker}, 
graphenes~\cite{Iorio:2011yz} 
and Berry phases~\cite{MartinMartinez:2010sg}. 
For other attempts, 
see \cite{Retzker2} and its reference.

In this study, as the issue of the Unruh effect in the BEC, 
we analyze the critical acceleration at which the BEC is formed 
based on the thermal excitation by the Unruh effect. 
Although many studies have been performed on 
the Unruh effect and the BEC until now, 
those have been usually performed separately, 
and the study such as ours is interesting.
\newline

We mention the critical acceleration we analyze in this study. 
We first consider an ideal gas 
in the Minkowski spacetime at zero temperature, 
where the gas is composed of particles 
described by a free complex scalar field at finite density. 
Since the gas is now at zero temperature, 
it is considered to be in the BEC state.   
%---
Then, we accelerate the ideal gas. 
At this time, 
since the gas will experience the Unruh temperature $T_U = \hbar\, a/(2\pi c \,k_B)$ 
by the Unruh effect, 
the BEC will be finally resolved 
at some point 
when gradually growing the acceleration. 
The critical acceleration we analyze in this study is this one.    
\newline

Next, we mention about the critical moment where the BEC is formed in our analysis.
We first analyze the effective potential in no BEC situation, 
then we obtain the particle density from that. 
%---
Then, from its representation, 
we find that 
if we decrease the acceleration 
fixing the particle density to a constant, 
either the particle's chemical potential 
or the value of the zero-mode of the field 
should grow. 
%---

In our analysis, 
we begin with the highly accelerating situation. 
Thus, there is no BEC, 
therefore the value of the zero-mode should be zero at this time, 
where its value means the expectation value of the BEC state. 
Then, we gradually decrease the acceleration 
keeping the particle density to a constant.  
Thus, the chemical potential grows. 
%---
However, 
we find that 
there is an upper bound 
for the value of the chemical potential. 
Thus, when the chemical potential reaches the upper bound, 
and if we further decrease the acceleration 
keeping the particle density to constant, 
the value of the zero-mode should begin to grow.

Thus, the moment that the chemical potential reaches the upper bound 
is the critical moment for the appearance of our BEC.  
\newline
  
The situation where our analysis is performed 
is the time just before the BEC is formed.  
More specifically, 
our analysis 
is performed 
in the situation that 
the values of the zero-mode is put to zero 
and the chemical potential is put to the upper bound.  
Then, we can obtain the critical acceleration from that. 

In our analysis, 
%due to some technical difficulty, 
we consider the situation $m/a_c \ll 1$, 
and take its leading contribution. 
Since the acceleration is proportional to the temperature in the representation the Unruh effect, 
we can regard this as 
$k_B T_c/2 \gg mc^2$~($T_c$ is the critical temperature given from $a_c$). 
Therefore, our analysis is in the relativistic situation 
where the energy of the particle's thermal motion is much higher than its static energy at the critical moment.
\newline

So far, several critical accelerations 
for the spontaneous symmetry breaking (SSB) 
by the Unruh effect have been analyzed~\cite{Ohsaku:2004rv, Ebert:2006bh,Castorina:2012yg}.
%---
Also, we introduce some interesting studies. 
%---
\cite{Hill:1985wi} argues that the Unruh effect does not contribute to the restoration 
of the SSB. 
%---
\cite{Benic:2015qha} argues that 
a larger acceleration enhances a condensate 
compared to those in a non-accelerated vacuum. 
%---
In \cite{Cavalcanti:2001jh}, 
although the background spacetime is not the Rindler space, 
whether the BEC can be formed or not is shown in an ideal bosonic gas model with a point-like impurity 
at finite temperature in some gravitational force in each of $D=1,2,3$. 
%---
It is shown in \cite{Takagi:1985vp} that, 
in odd dimensions,
a constantly accelerated detector detects 
free massless scalar particles 
as the one to follow the Fermi-Dirac distribution, 
and some solution for this problem is given in \cite{Ooguri:1985nv}.

%%%%%%%%%%%%%%%%%%%%%%%%%%%%%%%%%%%%%%%%%%%%%%%%%%%%%%%
\section{The model}
\label{ebwlr}
%%%%%%%%%%%%%%%%%%%%%%%%%%%%%%%%%%%%%%%%%%%%%%%%%%%%%%%

The model in this paper is a free complex scalar model 
at the finite density 
corresponding to an constantly accelerating ideal gas 
with an acceleration $a$. 
The Lagrangian density is given as
\begin{align}\label{action}
{\cal L}=  \, \hbar^2 g^{\mu\nu} \p_\mu \phi^* \p_\nu \phi - c^2 m^2 \phi^* \phi 
\end{align}
where $\phi\equiv \frac{1}{\sqrt{2}}(\phi_1+i \phi_2)$, 
and $\mu$, $\nu$ mean the Rindler coordinates 
which we explain below.

The background spacetime of a constantly accelerating system 
in the Minkowski spacetime 
can be given by the Rindler coordinate system.
its coordinates are given as follows: 
\begin{align}
(\eta,\rho,y,z) \equiv (\eta,\rho,x_\perp) \quad {\rm with} \quad x_\perp \equiv (y,z).  
\end{align}
This relates with the Minkowski coordinates $(t,x,x_\perp)$ as follows:
\begin{align}\label{Trans_to_Rindler}
(t,x) = \frac{c}{a} \big(  \sinh \frac{at}{c},\, c \cosh \frac{at}{c} \big) \equiv \rho \left(\sinh \eta,\, c \cosh \eta \right), 
\end{align}
where the accelerating direction has been thought to be in the $x$-direction.  
The Rindler metric \black{in our notation} is 
\begin{align}\label{Rindler_metric}
ds^2 = (c\rho)^2 d\eta^2 - d(c\rho)^2 - dx_\perp^2.  
\end{align}
In what follows, we use the unit system:~$c=\hbar=k_B=1$. 
\newline

The constantly accelerating one 
in the Minkowski spacetime 
corresponds to the one moving along a line 
on a constant $\rho$ in the Rindler coordinate system.  
The relation between $\rho$ and $a$ are $\rho=1/a$ 
as can be seen from (\ref{Trans_to_Rindler}). 
Since constantly accelerating one experiences 
the system as a thermal-bath with the Unruh temperature $T_U=a/2\pi$ by the Unruh effect, 
one moving along a line on a constant $\rho$ in the Rindler coordinates experiences the temperature 
\begin{align}\label{UnruhTemprerature}
T_U=1/2\pi \rho.
\end{align} 
Thus, the gas in our study is supposed 
to be in a thermal-bath with $T_U$ above. 
\newline

We can see from (\ref{UnruhTemprerature}) that 
varying $\rho$ means varying the temperature. 
For this reason, how to interpret the results is unclear, 
if the spacetime integrate including the $\rho$-integration were performed.    
This difficulty in the treatment of the coordinate $\rho$ is also mentioned 
at the chapter of conclusions and discussions 
in \cite{Iso:2010yq}. 
%in the context of the study of the Larmor radiation with the correction rooted in the Unruh effect.   

Let us turn to how this problem is treated
in other studies 
for the critical acceleration by the Unruh effect. 
%although it is difficult to explain it in here briefly. 
In \cite{Ohsaku:2004rv,Ebert:2006bh,Castorina:2012yg}, 
the action is given with the 4D spacetime integrate 
including the $\rho$-integration.
%, which point is the problem. 
However, their analysis is once performed 
using the Green's function 
that has dependence on the $\rho$-direction, 
then the $\rho$ is treated as a constant 
in the analysis of the effective potential. 
As a result, the $\rho$-integrate becomes 
just a volume factor 
in the analysis of the effective potential. 
%For more details, refer \cite{Ohsaku:2004rv,Ebert:2006bh,Castorina:2012yg}.  

In our analysis, 
as well as \cite{Ohsaku:2004rv,Ebert:2006bh,Castorina:2012yg}, 
we obtain the Green's function that 
has dependence on the $\rho$-direction. 
However, the spacetime integrate in our action is given 
without the $\rho$-integrate. 
As a result, $\rho$ is a parameter in our model.

At this time, we can see that 
the action is needed to be multiplied 
by a quantity with the dimension of length 
so that the action becomes dimensionless.  
For this purpose, 
we put $d\rho$ in our action, 
however the process 
to obtain the path-integral representation %given 
in (\ref{Z_DpDp}) 
%for the probability amplitude 
%from the operator formalism representation 
is supposed to have been performed fixing $\rho$. 
\newline

There are four regions separated by the event-horizons in the Rindler space. 
The region treated in this paper is the right wedge only.

%%%%%%%%%%%%%%%%%%%%%%%%%%%%%%%%%%%%%%%%%%%%%%%%%%%%%%%
\section{The effective potential}
%%%%%%%%%%%%%%%%%%%%%%%%%%%%%%%%%%%%%%%%%%%%%%%%%%%%%%%
%
%
%%%%%%%%%%%%%%%%%%%%%%%%%%%%%%%%%%%%%%%%%%%%%%%%%%%%%%%
\subsection{Performance of the path-integral}
%%%%%%%%%%%%%%%%%%%%%%%%%%%%%%%%%%%%%%%%%%%%%%%%%%%%%%%
%
% 
We begin with the probability amplitude:~
\begin{align}\label{Z_DpDp}
Z =& \int \! {{\cal D} \pi_\eta} {{\cal D} \phi} 
\exp \Big[ i \,  \int \! d^{3}x \, \blue{d\rho} \, \gamma \, 
\Big( \,\pi^{\eta}\p_\eta \phi + \,\pi^{\eta}{}^* \p_\eta \phi^* - \big({\cal H}-\mu \, q \big) \Big) \Big], 
\end{align}
where $\gamma \equiv \sqrt{-{\rm det}\,g_{\mu\nu}}$ and $d^3x \equiv d\eta \, d^2x\!_\perp$, 
and $\mu$ and $q$ are the chemical potential and the density of particles, respectively.
For the reason mentioned in Sec.\ref{ebwlr}, 
to fix the acceleration (Unruh temperature), 
the integrate of the $\rho$-direction is not included 
in the spacetime integrate in the action. 
%---
As a result, 
$\rho$ is a parameter in our analysis, 
and the Unruh temperature that the gas experiences is constant
according to the relation in (\ref{UnruhTemprerature}). 
%---
However, 
at this time, 
the action 
is needed to be multiplied 
by a quantity 
with the dimension of length 
so that the action can become dimensionless. 
%---
For this purpose, We put $d\rho$. 
However, the spacetime integrate is the one without the $\rho$-direction. 
Explanation for $\pi^{\eta}$ and ${\cal H}$ is given in the following.

$
\big(\pi^{\mu},\pi^{\mu}{}^*\big) 
\equiv 
\big( \frac{\p {\cal L}}{\p(\p_\mu \phi^*)},\,\frac{\p {\cal L}}{\p(\p_\mu \phi)} \big)
$ are the momenta given as
\begin{align}
\label{CanonicalMU}
\big(\pi^{\eta},\pi^{\eta}{}^*\big) 
&= 
\big( g^{\eta\eta} \p_\eta \phi^*,  \, g^{\eta\eta} \p_\eta \phi \big).  
\end{align}

${\cal H}$ is the Hamiltonian density given as
\begin{align}\label{Hamiltonian} 
{\cal H} 
&= \pi^{\eta} \,\p_\eta \phi + \pi^{\eta}{}^* \,\p_\eta \phi - {\cal L}\nn
&= \pi^{\eta}{}^* \pi_{\eta} - g^{ij} \p_i \phi^* \p_j \phi +m^2 \phi^* \phi,   
\end{align}
where $\pi^{\mu}$ is defined as 
$\pi^{\mu} \equiv \frac{1}{\sqrt{2}}(\pi^{\mu}_1+i \, \pi^{\mu}_2)$.  
Correspondingly, the functional integration measure of $\pi_\eta$ changes as 
${\cal D \pi_\eta} = {\cal D} \pi_{1\eta} {\cal D} \pi_{2\eta}$. 
From (\ref{CanonicalMU}), we can see
\begin{align}
\pi^{\eta}_{1,2} = g^{\eta\eta}\p_\eta \phi_{1,2}. 
\end{align}

It turns out that 
the conserved current associated with the U$(1)$ global symmetry 
in our model is obtained as 
\begin{align}
J^\mu &= -i \, g^{\mu\nu}(\phi\,\p_\nu\phi^*-\phi^*\,\p_\nu\phi).
\end{align}

Using this $J^\mu$, the integral of conserved charge density can be written as
\begin{align}\label{Density}   
\int \! d^{3}x \, \blue{d\rho} \, \gamma \, q 
= \int \! d^{3}x \, \blue{d\rho} \, \gamma \, J^\eta 
= \int \! d^{3}x \, \blue{d\rho} \, \gamma \, \big(-\pi^{\eta}_2 \, \phi_1 + \pi^{\eta}_1 \, \phi_1 \big). 
\end{align}
 
Substituting (\ref{Hamiltonian}) and (\ref{Density}) into (\ref{Z_DpDp}), we can obtain the following $Z$:
\begin{align}
Z 
&= \int \! {\cal D} \pi_{1\eta} {\cal D} \pi_{2\eta}{\cal D \phi} \exp \Big[ 
\frac{i}{2} \int \! d^{3}x \, \blue{d\rho} \, \gamma \, \Big\{ 
\nn
%---
& \quad
-\frac{1}{2}\big(\pi^{\eta}_1\pi_{1\eta}-2\,(\p_\eta \phi_1 + \mu \,\phi_2)\,\pi^{\eta}_1 \big)
%---
-\frac{1}{2}\big(\pi^{\eta}_2\pi_{2\eta}-2\,(\p_\eta \phi_2 + \mu \,\phi_1)\,\pi^{\eta}_2 \big)
\nn
%---
& \quad
+\frac{1}{2}\big( (\p_i \phi_1)^2+(\p_i \phi_2)^2-m^2(\phi_1^2+\phi_2^2) \big) 
\Big\} \Big], 
\end{align}
where $i,j=x_\perp$. 
Here, in the above, we perform the following rewritings:
\begin{subequations}
\begin{align}
 & \pi^{\eta}_1\pi_{1\eta}-2\,(\p_\eta \phi_1 + \mu \,\phi_2)\,\pi^{\eta}_1
=\,\, g^{\eta\eta} \Big\{\big( \pi_{1\eta} - (\p_\eta \phi_1 + \mu \,\phi_2) \big)^2 - (\p_\eta \phi_1 + \mu \,\phi_2)^2 \Big\}, \\
%=====
 & \pi^{\eta}_2\pi_{2\eta}-2\,(\p_\eta \phi_2 + \mu \,\phi_1)\,\pi^{\eta}_2
=\,\, g^{\eta\eta} \Big\{\big( \pi_{2\eta} - (\p_\eta \phi_2 + \mu \,\phi_1) \big)^2 - (\p_\eta \phi_2 + \mu \,\phi_1)^2 \Big\}.
\end{align}
\end{subequations}
Furthermore, we redefine the fields as
\begin{subequations}
\begin{align}
& \pi_{1\eta} - (\p_\eta \phi_1 + \mu \,\phi_2) \rightarrow \pi_{1\eta},\\
& \pi_{2\eta} - (\p_\eta \phi_2 + \mu \,\phi_1) \rightarrow \pi_{2\eta}.
\end{align}
\end{subequations}
As a result, we can write $Z$ as
\begin{align}
\label{Z_Dp}
Z 
= & \,\, 
{\cal C} \int \! {{\cal D} \phi} \exp \Big[ 
\frac{i}{2} \int \! d^{3}x \, \blue{d\rho} \, \gamma \, \Big( g^{\eta\eta}(\p_\eta\phi_1+\mu \,\phi_2)^2 + (\p_i \phi_1)^2 
\nn
%---
& \qquad \qquad \qquad \qquad \quad \,\,
+ \, g^{\eta\eta}(\p_\eta\phi_2+\mu \,\phi_1)^2 + (\p_i \phi_2)^2  - m^2 ({\phi_1}^2+{\phi_2}^2)\Big) \Big]. 
\end{align}
with 
$\displaystyle 
{\cal C} \equiv \int \! {\cal D} \pi_{1\eta} {\cal D} \pi_{2\eta}  
\exp \Big[ i \int \! d^{3}x \, \blue{d\rho} \, \gamma \, g^{\eta\eta} \big( (\pi_{1\eta})^2+(\pi_{2\eta})^2 \big) \Big].  
$
Performing the path-integral of $\pi_{1\eta}$ and $\pi_{2\eta}$ formally, 
we think that ${\cal C}$ become some factor. We ignore ${\cal C}$ in what follows. 
As a result, with some straight forward calculation, we can write $Z$ in the following form,
\begin{align}
\label{Z_Dp}
Z 
= & \, 
\int \! {\cal D \phi} \exp 
\Big[ \! -\frac{i}{2} \int \! d^{3}x \, \blue{d\rho} \,\gamma \, \Big( \phi_1 G \phi_1 + \phi_2 G \phi_2
%---
+ 2g^{\eta\eta}\mu \, (\phi_2 \p_\eta \phi_1 - \phi_1 \p_\eta \phi_2 ) \Big) \Big],
\end{align}
where $G \equiv \p_\eta^2+ \gamma^{-1}g^{ij}\p_i(\gamma\p_j)+m^2-g^{\eta\eta}\mu^2$. 
 
Now, let us rewrite 
the real and imaginary parts of the field 
into a convenient expression 
for the analysis of the Bose-Einstein condensation (BEC).   
As we have written in Sec.\ref{vnrver}, 
the BEC can be considered 
as the situation that all particles get into the least energy state. 
The least energy state can be considered as the zero-mode of the field, 
and the situation that all the particles are in the least 
energy state can be considered 
as the condensation of the zero-mode. 
Thus, a convenient expression for the analysis of the 
BEC in our analysis is the one in which the zero-mode is separated as
\begin{subequations}\label{phi_def}
\begin{align} 
\phi_1 \, \equiv& \, \sqrt{2} \, \alpha \, \cos \theta + \hat{\phi}_1, \label{BEC_representation1}\\
\phi_2 \, \equiv& \, \sqrt{2} \, \alpha \, \sin \theta + \hat{\phi}_2, \label{BEC_representation2}
\end{align}
\end{subequations}
where $\alpha$ represents the value of the zero-mode's condensation, 
$\theta$ is a phase in the zero-mode, 
and $\hat{\phi}_1$, and $\hat{\phi}_2$ are the non-zero modes.
%---
Depending on just before or after the condensation starts, $\alpha$ behaves as follows:
\begin{align}
&\quad \alpha     = 0 ~:~ \textrm{before the condensation} \nn
&\quad \alpha \not= 0 ~:~ \textrm{after the condensation} \nonumber
\end{align}

At this time, $\phi_1 G \,\phi_1$ and $\phi_2 G \,\phi_2$ can be written as
\begin{align} 
\phi_1 G \,\phi_1 =& \,\, 2\alpha^2(m^2-g^{\eta\eta}\mu^2)\,\cos^2 \theta+\hat{\phi}_1 G \,\hat{\phi}_1, \\
\phi_2 G \,\phi_2 =& \,\, 2\alpha^2(m^2-g^{\eta\eta}\mu^2)\,\sin^2 \theta+\hat{\phi}_2 G \,\hat{\phi}_2.
\end{align}

As a result, we can rewrite $Z$ into 
\begin{align}  
& Z =   
\, \exp\Big[ \! -i \, \alpha^2 \! \int \! d^{3}x \, \blue{d\rho} \,\gamma \, (m^2-g^{\eta\eta}\mu^2)\Big] 
\int\!  {\cal D \hat{\phi}} 
\exp \bigg[
\nn   
%-----
& \qquad
- \frac{i}{2} \int \! d^{3}x \, \blue{d\rho}
\, \gamma \left( \! \begin{array}{cc} \hat{\phi}_1 & \hat{\phi}_2 \end{array} \! \right) \! 
\left( \! \begin{array}{cc} G & -2g^{\eta\eta}\mu \,\p_\eta \\ 2g^{\eta\eta}\mu \,\p_\eta & G \end{array} \! \right) \!
\left( \! \begin{array}{c} \hat{\phi}_1 \\ \hat{\phi}_2 \end{array} \! \right) \nn
%-----
& \qquad + \sqrt{2} \, \alpha \cos \theta \, \Big( \, G \, (\phi_1+\phi_2) + (\phi_1+\phi_2) \, G \, \Big)  \bigg].
\end{align}

Our analysis is performed just before the condensation as mentioned later.  
For this reason, we put $\alpha=0$ in what follows. 
As a result, we can write $Z$ as
\begin{align}
Z = \int\!  {\cal D \hat{\phi}} \exp \bigg[ 
%-----
- \frac{i}{2} \int \! d^{3}x \, \blue{d\rho}
\, \gamma \left( \! \begin{array}{cc} \hat{\phi}_1 & \hat{\phi}_2 \end{array} \! \right) \! 
\left( \! \begin{array}{cc} G & -2g^{\eta\eta}\mu \,\p_\eta \\ 2g^{\eta\eta}\mu \,\p_\eta & G \end{array} \! \right) \!
\left( \! \begin{array}{c} \hat{\phi}_1 \\ \hat{\phi}_2 \end{array} \! \right)  
\bigg].
\end{align}
Then, we can see from the form of $G$ given below (\ref{Z_Dp}) that there should be the condition: 
\begin{align}\label{bound_of_mu}
m^2-g^{\eta\eta}\mu^2 \ge 0,
\end{align} 
otherwise the path-integrals of $Z$ diverges at the configuration that all the momenta are zero. 
This relation gives the upper limit of the chemical potential for a given mass and an Unruh temperature. 

Performing the diagonalization as
\begin{align}\label{vwhvl}
Z 
& \,
= \int\! {\cal D} \hat{\phi} \exp \bigg[ - \frac{i}{2} \int \! d^{3}x \, \blue{d\rho} \, \gamma  
%-----
\left( \! \begin{array}{cc} \hat{\phi}_1 & \hat{\phi}_2 \end{array} \! \right) UU^{-1}
\left( \! \begin{array}{cc} G & -2g^{\eta\eta}\mu \,\p_\eta \\ 2g^{\eta\eta}\mu \,\p_\eta & G \end{array} \! \right) 
UU^{-1} \left( \! \begin{array}{c} \hat{\phi}_1 \\ \hat{\phi}_2 \end{array} \! \right)  
\bigg].
\nn 
%=====
& 
= \int\!  {\cal D' \hat{\phi'}} \exp \bigg[ - \frac{i}{2} \int \! d^{3}x \, \blue{d\rho} \, \gamma   
%-----
\left( \! \begin{array}{cc} \hat{\phi}'_1 & \hat{\phi}'_2 \end{array} \! \right) \! 
\left( \! \begin{array}{cc} G + 2g^{\eta\eta}\mu \,\p_\eta & 0 \\ 0 & G - 2g^{\eta\eta}\mu \,\p_\eta \end{array} \! \right) 
\!
\left( \! \begin{array}{c} \hat{\phi}'_1 \\ \hat{\phi}'_2 \end{array} \! \right)  
\bigg],
\end{align}
where $U \equiv \frac{1}{\sqrt{2}} \left( \begin{array}{cc} i &  -i \\ 1 & 1 \end{array} \right)$ is a unitary matrix defined 
to perform the above diagonalization, and correspondingly 
$\left( \! \begin{array}{c} \hat{\phi}'_1 \\ \hat{\phi}'_2 \end{array} \! \right) \equiv U^{-1} \left( \! \begin{array}{c} \hat{\phi}_1 \\ \hat{\phi}_2 \end{array} \! \right)$. 
At this transformation, the functional measure is also transformed, which we have described as ${\cal D \hat{\phi}} \rightarrow {\cal D' \hat{\phi'}}$. 
However, since $U$ is a constant unitary matrix, the difference between ${\cal D \hat{\phi}}$ and ${\cal D' \hat{\phi'}}$ contribute only to some 
constant factor in the path-integral, and we ignore it in what follows. 

Performing the path-integral, we can obtain   
\begin{align}\label{one-loop}
Z \,=& \,
\Det \Big(\big( G + 2g^{\eta\eta}\mu \,\p_\eta \big)\big( G - 2g^{\eta\eta}\mu \,\p_\eta \big)\Big)^{-1/2}.
\end{align}
In the above, with regard to the treatment of $d\rho$, we consider that the integration of this has been performed by assigning a value at one point.  
Hence, the effective action $W$ defined as $Z=\exp i W$ can be written as 
\begin{align}
W
%-----
\,=& \,
\frac{i}{2} \, 
\black{\Log\,\Det} \Big(\big( G + 2g^{\eta\eta}\mu \,\p_\eta \big)\big( G - 2g^{\eta\eta}\mu \,\p_\eta \big)\black{\Big)}
\nn
%-----  
=& \,
\frac{i}{2} \,  
\black{\Tr\,\Log} \Big(\big( G + 2g^{\eta\eta}\mu \,\p_\eta \big)\big( G - 2g^{\eta\eta}\mu \,\p_\eta \big)\black{\Big)}
\nn
%=====
=& \,
\frac{i}{2} \, V \! \int \! \frac{dk^3}{(2\pi)^3} 
\Big(\,\,\,
\Log \big( \p_\eta^2+ \gamma^{-1}g^{ij}\p_i(\gamma\,\p_j)+M^2+2ig^{\eta\eta}\mu\,\p_\eta \big) 
\nn
%-----
\label{W01} % <--- LABEL
& \qquad \qquad \quad \,
+ 
\Log \big( \p_\eta^2+ \gamma^{-1}g^{ij}\p_i(\gamma\,\p_j)+M^2-2ig^{\eta\eta}\mu\,\p_\eta \big)
\Big),
\end{align}
where $M^2 \equiv m^2-g^{\eta\eta}\mu^2$ and $V \equiv \int d^3x \, \gamma$ is the volume for the $(\eta, x_\perp)$ spacetime, 
which appears from \black{the rewriting of} a functional trace into an integration:  
\begin{align}\label{TrToInt}
\black{\Tr 
\rightarrow \left(\frac{L}{2\pi}\right)^3 \left(\frac{2\pi}{L} \right)^3 \sum_k
\equiv
V \int \frac{dk^3}{(2\pi)^3}} 
\end{align} 
with 
$\left( \frac{2\pi}{L} \right)^3 \sum_k = \int dk^3$ 
and $V \equiv L^3 = \int d^3x \, \gamma$, 
where $L$ means just length in each space for the $\eta$, $x_\perp$-directions, 
and $\sqrt{-{\rm det}\,g_{ab}} = \sqrt{-g_{\rho\rho}\,{\rm det}\,g_{ab}} = \gamma$~($a,b=\eta,x_\perp$ except for $\rho$, and $g_{\rho\rho}=1$).   
$d^3k \equiv d\omega \, d^2k\!_\perp$ with $k\!_\perp \equiv (k_y, k_z)$, 
which are momenta corresponding to the coordinate $(\eta,x_\perp )$ as in (\ref{Fourier_trans}). 

Defining $\delta_\pm \equiv g^{\eta\eta} \p_\eta^2+ \gamma^{-1}g^{ij}\p_i(\gamma\,\p_j) \pm 2ig^{\eta\eta}\mu\,\p_\eta$, 
we further rewrite $W$ as
\begin{align}
W
=& \,
\frac{i}{2} \, V \! \int \! \frac{dk^3}{(2\pi)^3} 
\Big(\, \Log \big( \delta_+ + M^2 \big) + \Log \big( \delta_- + M^2 \big) \Big)
\nn
%=====
=& \,
\frac{i}{2} \, V \! \int \! \frac{dk^3}{(2\pi)^3} 
\,\bigg(\,\,
\int_0^{M^2}\!\!d\Delta^2(\delta_+ + \Delta^2)^{-1} + \Log \, \delta_+  
%-----
\label{W02} 
+ \int_0^{M^2}\!\!d\Delta^2(\delta_- + \Delta^2)^{-1} + \Log \, \delta_-\bigg).  
\end{align}

Here, we can ignore $\int \frac{dk^3}{(2\pi)^3} \, \Log\,\delta_\pm$ 
for the reason in what follows:~First, since we perform the derivative with regard to the chemical potential to obtain the particle density at the end, 
we look at only the part concerning the chemical potential as
\begin{align}
& 
\int\! \frac{dk^3}{(2\pi)^3}\,\,\Log \, \delta_\pm
\nn
%---
\black{=} & \black{\int\! \frac{dk^3}{(2\pi)^3} \,\Log \,\, \big( \p_\eta^2+ \gamma^{-1}g^{ij}\p_i(\gamma\,\p_j) \big)}
%---
\black{+ \int\! dk_\eta \frac{dk_\perp^2}{(2\pi)^3} \,\Log \left( 1 \pm \frac{2ig^{\eta\eta}\mu\,\p_\eta}{\p_\eta^2 + \gamma^{-1}g^{ij}\,\p_i(\gamma\,\p_j)} \right)}
\nn
%---
\black{\sim} & \black{\int\! dk_\eta \frac{dk_\perp^2}{(2\pi)^3} \,\Log \left( 1 \pm \frac{2ig^{\eta\eta}\mu\,\p_\eta}{\p_\eta^2 + \gamma^{-1}g^{ij}\,\p_i(\gamma\,\p_j)} \right).}     
\end{align} 
In the one above, we have described the integral $\int dk^3$ separately as $\int dk_\eta \int dk^2_\perp$. 
Then, replacing $\p_{i}$ with $i\,k_{i}$, from the fact that $\int dk \, k \log\,(1+\frac{c_1k }{c_2+k^2})=0$~($c_{1,2}$ are some constants), 
we can see that the $\mu$-dependent part vanishes in the $k_\perp$-integrals. 

Finally, we can write the effective action $W$ in (\ref{W02}) as
\begin{align}\label{W_Dpm}
W = \frac{i}{2}V \, \gamma \! \int \! \frac{dk^3}{(2\pi)^3} \int^{M^2}_0 \!\!\! d\Delta^2 \, (\widetilde{D}_+ + \widetilde{D}_-),
\end{align}
where $\widetilde{D}_\pm$ is 
\begin{align}\label{def_D}
\blue{\gamma^{-1}}
\left(\delta_\pm + \Delta^2 \right)^{-1} 
\equiv
\widetilde{D}_\pm 
=
\widetilde{D}_\pm(k_\eta,\rho,k_\perp). 
\end{align} 
In the one above, we wrote the arguments concerning the momenta as $k_\eta$ and $k_\perp$ despite that these are given 
by the differential operators in the actual expression. %, which may be allowed.

%%%%%%%%%%%%%%%%%%%%%%%%%%%%%%%%%%%%%%%%%%%%%%%%%%%%%%%
\subsection{Analysis of $\widetilde{D}_\pm$}
%%%%%%%%%%%%%%%%%%%%%%%%%%%%%%%%%%%%%%%%%%%%%%%%%%%%%%%

From the definition in (\ref{def_D}), we can have the following equation:  
\begin{align}\label{Identity_of_Dpm}
\delta^4(x-x') =& \blue{\gamma}\left(\delta_\pm + \Delta^2 \right) D_\pm(x-x') \nn
               =& \blue{\gamma}\left(g^{\eta\eta}\partial_\eta^2+ \gamma^{-1}g^{ij}\partial_i(\gamma\p_j)\pm 2i\,g^{\eta\eta}\mu\,\partial_\eta+\Delta^2 \right) D_\pm(x-x')  
\end{align}
with $D_\pm(x-x')$ defined as
\begin{align}\label{Fourier_trans}
D_\pm(x-x') = \int \! \frac{d^3k}{(2\pi)^3} \, \widetilde{D}_\pm \, e^{i\,\big(\omega \, (\eta-\eta') - k\!_\perp (x\!_\perp-x\!_\perp')\big)}.  
\end{align} 
We can rewrite (\ref{Identity_of_Dpm}) as 
\begin{align}
&\delta(\rho-\rho') \int \! \frac{d^3k}{(2\pi)^3} \, e^{i\,\big(\omega \, (\eta-\eta') - k\!_\perp (x\!_\perp-x\!_\perp')\big)} \nn
%-----
=& \,\, \int \! \frac{d^3k}{(2\pi)^3} \,\blue{\gamma} 
\left( g^{\eta\eta}\p_\eta^2+ \gamma^{-1}g^{ij}\p_i(\gamma\p_j)\right. \left. \pm 2i\,g^{\eta\eta}\mu\,\p_\eta+\Delta^2 \right)
\widetilde{D}_\pm \, e^{i\,\big(\omega \, (\eta-\eta') - k\!_\perp (x\!_\perp-x\!_\perp')\big)}  \nn
%=====
=& \,\, \int \! \frac{d^3k}{(2\pi)^3} \,\blue{\gamma}
\left( \rho^{-2}\p_\eta^2 - \rho^{-1}\p_\rho(\rho\p_\rho) - (\p_x^2+\p_y^2)\right.
%-----
\label{GreenFuncEOM2}
\left. 
\pm 2i\,g^{\eta\eta}\mu\,\p_\eta+\Delta^2 \right)
\widetilde{D}_\pm \, e^{i\,\big(\omega \, (\eta-\eta') - k\!_\perp (x\!_\perp-x\!_\perp')\big)}, 
\end{align}  
where $d^3k \equiv d \omega \, dk_\perp^2$. As a result, we can obtain the equation which $\widetilde{D}_\pm$ should satisfy as
\begin{align}
\blue{\gamma^{-1}} \delta(\rho-\rho') &=\big( 
-\rho^{-2}\omega^2 
-\rho^{-1}\p_\rho - \p_\rho^2 + k_y^2+k_z^2
%---
\pm 2\rho^{-2} \mu \, \omega 
+ \Delta^2 \, \big) \widetilde{D}_\pm. 
\end{align}
Finally, we can obtain the equation that determines $\widetilde{D}_\pm$ as
\begin{align}\label{GreenFuncEOM2}
\big(\,{\cal F} + \Omega_\pm^2\,\big) \, \widetilde{D}_\pm &= -\blue{\rho} \, \delta(\rho-\rho'), 
\end{align}
where ${\cal F} \equiv \rho^2 \p_\rho^2+ \rho  \, \p_\rho - \rho^2 \kappa^2$ with 
$\kappa^2\equiv k^2+\Delta^2$, ($k^2 \equiv k_y^2+k_z^2$), and $\Omega_\pm ^2 \equiv \omega (\omega \mp 2\mu)$.  

Let us now determine $\widetilde{D}_\pm$ from (\ref{GreenFuncEOM2}). 
First, we can see that ${\cal F}$ can satisfy the following differential equation:
\begin{align}\label{Eigen_equation_of_F}
{\cal F} \,\Theta_\lambda(\rho,k) = (i \lambda)^2 \, \Theta_\lambda(\rho,k). 
\end{align}
with
\begin{align}
\Theta_\lambda(\rho,k)= C_\lambda K_{i \lambda}(\kappa \rho),\quad C_\lambda = \frac{1}{\pi}\sqrt{2\lambda \sinh(\pi \lambda)}, 
\end{align}
where the values of $\lambda$ are positive real numbers, 
$K_{i \lambda}(\kappa \rho)$ means the second kind modified Bessel function, 
and $C_\lambda$ was obtained from 
\begin{align}\label{Normalization_condition_of_Theta}
\int_0^\infty \frac{d\rho}{\rho} \, \Theta_{\lambda'}(\rho,k) \, \Theta_\lambda(\rho,k)=\delta(\lambda'-\lambda). 
\end{align} 

We then assume that $\widetilde{D}_\pm$ can be written 
by taking $\Theta_\lambda(\rho,k)$ as the bases 
for the orthogonal directions labeled by $\lambda$ as 
\begin{align}\label{Theta_expansion_of_D}
\widetilde{D}_\pm=\int_0^\infty \! d\lambda \, f_{\lambda,\pm}(\omega,\rho') \, \Theta_\lambda(\rho,k),  
\end{align}
where the above means an orthogonal base expansion of $\Theta_\lambda(\rho,k)$ labeled by $\lambda$, 
and $f_{\lambda,\pm}(\omega,\rho')$ are the coefficients of each direction. 
If we can find $f_{\lambda,\pm}(\omega,\rho')$ that satisfies (\ref{GreenFuncEOM2}), this assumption is right.  
We now obtain such $f_{\lambda,\pm}(\omega,\rho')$.

We can see that now we can have the following two independent equations as 
\begin{subequations}
\begin{align} 
%--- 
\label{DFTheta}
&\int \frac{d\rho}{\blue{\rho}} \, \widetilde{D}_\pm \, {\cal F} \, \Theta_\lambda(\rho,k)
= -\lambda^2 \int \frac{d\rho}{\blue{\rho}} \, \widetilde{D}_\pm \, \Theta_\lambda(\rho,k), \\
%---
\label{ThetaFD}
&\int \frac{d\rho}{\blue{\rho}} \, \Theta_\lambda(\rho,k) \, {\cal F} \widetilde{D}_\pm
= -\Omega_\pm^2 \int \frac{d\rho}{\blue{\rho}} \, \Theta_\lambda(\rho,k) \, \widetilde{D}_\pm 
- \Theta_\lambda(\rho',k).
%---
\end{align}
\end{subequations} 
Here, in obtaining (\ref{DFTheta}) and (\ref{ThetaFD}), 
we have used (\ref{Eigen_equation_of_F}) and (\ref{GreenFuncEOM2}), respectively. 

(\ref{DFTheta}) and (\ref{ThetaFD}) are equivalent to each other. 
Actually, this equivalence can be seen easily 
by what (\ref{DFTheta}) and (\ref{ThetaFD}) can be represented as 
$\big\langle \widetilde{D}_\pm \big| {\cal F} \big| \Theta_\lambda(\rho,k) \big\rangle$ 
and 
$\big\langle \Theta_\lambda(\rho,k) \big| {\cal F} \big| \widetilde{D}_\pm \big\rangle$, respectively. 
Then, by subtracting these two equations each other, 
\begin{align} 
0 &= \big(\! -\lambda^2+\Omega_\pm^2 \big) \int \frac{d\rho}{\blue{\rho}} \widetilde{D}_\pm \, \Theta_\lambda(\rho,k) + \Theta_\lambda(\rho',k)\nn
%=====
&= \big(\! -\lambda^2+\Omega_\pm^2 \big) \int \frac{d\rho}{\blue{\rho}} \cdot \int \!d\lambda' \, f_{\lambda',\pm}(\omega,\rho') \Theta_{\lambda'}(\rho,k) \cdot \Theta_\lambda(\rho,k) 
+ \Theta_\lambda(\rho',k)\nn
%=====
&= \big(\! -\lambda^2+\Omega_\pm^2 \big)
\int \!d\lambda' \cdot 
\int \frac{d\rho}{\blue{\rho}}\,  \Theta_{\lambda'}(\rho,k)\Theta_\lambda(\rho,k) \cdot 
f_{\lambda',\pm}(\omega,\rho')  
+ \Theta_\lambda(\rho',k)\nn
%=====
&= \big(\! -\lambda^2+\Omega_\pm^2 \big)
\int \!d\lambda' \cdot 
\delta(\lambda'-\lambda) \cdot 
f_{\lambda',\pm}(\omega,\rho') + \Theta_\lambda(\rho',k)\nn
%=====
&= \big(\! -\lambda^2+\Omega_\pm^2 \big) \, f_{\lambda,\pm}(\omega,\rho') + \Theta_\lambda(\rho',k),
\end{align}
where we have used (\ref{Normalization_condition_of_Theta}) and (\ref{Theta_expansion_of_D}).  
From the above, $f_{\lambda,\pm}(\omega,\rho')$ can be determined as 
\begin{align}\label{f_result}
f_{\lambda,\pm}(\omega,\rho') 
= \frac{\Theta_\lambda(\rho',k)}{\lambda^2-\Omega_\pm^2}. 
\equiv \frac{\Theta_\lambda(\rho',k)}{\lambda^2-\Omega^2}.
\end{align}
In the above, we have regarded $\Omega_+^2$ and $\Omega_-^2$ as the same, 
and denoted these as $\Omega^2$.
This is because the integrations of $f_{\lambda,+}(\omega,\rho')$ and $f_{\lambda,-}(\omega,\rho')$ with regard to $\omega$ in (\ref{Fourier_trans})
do not produce any difference under the transformation $\omega \rightarrow -\omega$. Namely, 
\begin{align}\label{cbwec}
D_+(x-x') 
%====================
&= \int \! \frac{d\omega \, d^2k_\perp}{(2\pi)^3} \, \cdot 
\int \! d\lambda \, \frac{\Theta_\lambda(\rho',k)\, \Theta_\lambda(\rho,k)}{\lambda^2-\omega (\omega - 2\mu)} \,
e^{i\,\big(\omega \, (\eta-\eta') - k\!_\perp (x\!_\perp-x\!_\perp')\big)}\nn
%====================
&= \int \! \frac{d\omega \, d^2k_\perp}{(2\pi)^3} \, \cdot 
\int \! d\lambda \, \frac{\Theta_\lambda(\rho',k)\, \Theta_\lambda(\rho,k)}{\lambda^2-\omega (\omega + 2\mu)} \,
e^{i\,\big(\red{-}\omega \, (\eta-\eta') - k\!_\perp (x\!_\perp-x\!_\perp')\big)}\nn
%====================
 &= D_-(x-x'). 
\end{align}
As a result, we can put $\Omega_\pm^2 \equiv \Omega^2$. 
Correspondingly, there is no difference between $\widetilde{D}_+$ and $\widetilde{D}_-$.
Hence, we put $\widetilde{D}_\pm \equiv \widetilde{D}$ and $D_\pm \equiv D$. 

Now, substituting (\ref{f_result}) into (\ref{Theta_expansion_of_D}), we can write $\widetilde{D}$ as
\begin{align}
\widetilde{D} 
&= \int \! d\lambda \,\frac{\Theta_\lambda(\rho',k)\Theta_\lambda(\rho,k)}{\lambda^2-\Omega^2}. 
\end{align}
\newline

Finally, the effective action $W$ can be written as
\begin{align}\label{W_Dpm2}
W = i\, V \blue{\gamma} \! \int \! \frac{dk^3}{(2\pi)^3} \int^{M^2}_0 \!\!\! d\Delta^2 \, \widetilde{D}.
\end{align}

%%%%%%%%%%%%%%%%%%%%%%%%%%%%%%%%%%%%%%%%%%%%%%%%%%%%%%%
\subsection{Analysis with the Euclidianization}
%%%%%%%%%%%%%%%%%%%%%%%%%%%%%%%%%%%%%%%%%%%%%%%%%%%%%%%

Now that $\widetilde{D}_\pm=\widetilde{D}$ have been obtained, backing to (\ref{W_Dpm}), 
we perform the integrates. 
We begin with performing the Wick rotation 
toward the $\eta$-direction. 
At this time, 
the metric given in (\ref{Rindler_metric})  
changes to
\begin{align} 
ds_E^2=\rho^2d\eta^2+d\rho^2+dx\!_\perp^2. 
\end{align}
From this, 
we can see that 
the $\eta$-direction is $S^1$-compactified 
with the period $\beta=2\pi$. 

Then, the following replacement arises: 
\begin{align} 
V \equiv \int d^3x \, \gamma \rightarrow -i \beta \int d^2x\!_\perp \, \gamma_E \equiv -i \beta \, V_E
\end{align}
with  
\begin{align}
\gamma &\equiv \sqrt{-g} \rightarrow \gamma_E \equiv \sqrt{g} = \rho, \\
\label{oive}
\int d \omega &\rightarrow \frac{2\pi i}{\beta} \sum_n 
\quad \textrm{and} \quad
\omega \rightarrow \omega_n = n,  
\end{align}  
where $V$ is defined under (\ref{TrToInt}), 
and now $\omega_n = 2 \pi n/R$ ($R=2\pi$). 
Since the values of $\Omega$ are discretized, 
we change the notation of $\Omega$ as $\Omega \rightarrow \Omega_{n}$. 

As a result, the effective potential $\Gamma$ defined as $W_E \equiv \beta \, V_E   \, \Gamma$ with $\exp(iW) \equiv \exp(-W_E)$ 
can be obtained as
\begin{align}\label{Gamma01}
\Gamma=& \,\,
\frac{\blue{\rho}}{\pi^3} 
\sum_{n=-\infty}^\infty \int \! \frac{d^2k\!_\perp}{(2\pi)^2} 
\int^{M^2}_0 \!\!\! d\Delta^2 \int \! d\lambda \, \black{\lambda} \, \sinh(\pi \lambda) \,
\frac{K_{i\lambda}^2(\kappa\blue{\rho})}{\lambda^2-\Omega_{n}^2} 
\nn
%---
=& \,\,
\frac{\blue{\rho}}{2\pi^3} \int \! d \lambda \, \black{\lambda} \, \sinh(\pi \lambda) \, \Phi \! \int \! dk \, k \Psi.
\end{align}
In the above, $d^2k\!_\perp=2\pi k dk$~($0 \le k \le \infty$),  
\begin{align}\label{Psi_def}
\Psi \equiv \int^{M^2}_0 \!\! d\Delta^2 \, K_{i\lambda}^2(\kappa\blue{\rho}),
\end{align}
and we have non-perturbatively evaluated the summation of $n$ as 
\begin{align}\label{summation-n}
\sum^{\infty}_{n=-\infty} \frac{1}{\lambda^2-\Omega_{n}^2} 
%---
=& \,\,
\frac{\pi}{\red{2}\sqrt{\lambda^2+\mu^2}}
%---
\Big(
\coth \big(\pi (\mu-\sqrt{\lambda^2+\mu^2}) \big) 
- 
\coth \big(\pi (\mu+\sqrt{\lambda^2+\mu^2}) \big) 
\Big)
\nn 
%---
\equiv& \,\, \pi \, \Phi,
\end{align}  
where $\Omega_{n}^2$ is now given as $ n (n \mp 2\mu)$, 
however no difference for $\pm$ appears in the r.h.s. 
 
Since we finally calculate the critical acceleration of the BEC, 
we focus on the critical moment, 
in which the value of the chemical potential is given by $m/a_c$, 
where $a_c$ means the critical acceleration. 
We explain this in what follows.

%%%%%%%%%%%%%%%%%%%%%%%%%%%%%%%%%%%%%%%%%%%%%%%%%%%%%%%
\subsection{The chemical potential at the critical moment}
\label{bvqdv}
%%%%%%%%%%%%%%%%%%%%%%%%%%%%%%%%%%%%%%%%%%%%%%%%%%%%%%%

Using the effective potential $\Gamma$ given in (\ref{Gamma01}), the particle density $d=-\p \Gamma/\p \mu$ can be written as
\begin{align}\label{density1}
d = & \, 
- \frac{\blue{\rho}}{2\pi^3} \int \! d\lambda \, \black{\lambda} \, \sinh(\pi \lambda) 
\bigg(\frac{\p \Phi}{\p \mu} \int \! dk k \, \Psi + \Phi \int \! dk k \, \frac{\p \Psi}{\p \mu} \bigg). 
\end{align}
Here, to perform $\frac{\p \Psi}{\p \mu}$, we show the $\mu$-dependence of $\Psi$ by expanded it around $\mu=0$ as
\begin{align}
\Psi =& \, \Psi\Big|_{\mu=0} 
+            \frac{\p \Psi}{\p \mu}     \bigg|_{\mu=0} \mu 
+ \frac{1}{2}\frac{\p^2 \Psi}{\p \mu^2} \bigg|_{\mu=0} \mu^2 
+ \cdots \nn
%=====
=& \int_0^{m^2} \!\! d \Delta \, K_{i\lambda}^2(\kappa\blue{\rho}) - a^2 K_{i\lambda}^2(\kappa\blue{\rho}) \Big|_{\Delta^2=m^2} \, \mu^2 + \cdots
\end{align}
with
\begin{subequations}
\begin{align}
\frac{\p \Psi}{\p \mu} \bigg|_{\mu=0} 
&= 
\frac{\p M^2}{\p \mu} \frac{\p \Psi}{\p M^2} \bigg|_{\mu=0} = 0,\\
%============
\frac{1}{2}\frac{\p^2 \Psi}{\p \mu^2} \bigg|_{\mu=0}
&=
\frac{1}{2}\frac{\p }{\p \mu} \bigg(\frac{\p M^2}{\p \mu} \frac{\p \Psi}{\p M^2} \,\bigg)\bigg|_{\mu=0}
=
-a^2 \frac{\p \Psi}{\p M^2} \bigg|_{\Delta^2=m^2}.
\end{align}
\end{subequations}
Using the above expansion, we can write the particle density as 
\begin{align}\label{density2}
d =& - \frac{\blue{\rho}}{2\pi^3} \int \! d\lambda \, \black{\lambda} \, \sinh(\pi \lambda) 
%\nn
%& 
\bigg(\frac{\p \Phi}{\p \mu} \int \! dk k \, \Psi - 2 a^2\mu \, \Phi K_{i\lambda}^2(\kappa\rho)\Big|_{\black{\Delta^2=m^2}} \bigg),
\end{align}
 
We can always confirm numerically 
that $\frac{\p \Phi}{\p \mu}<0$ and $\Phi>0$. 
Further, we can see $\Psi>0$, 
since the integrand in $\Psi$ is given by the square 
and the integral direction is positive, 
as can be seen in (\ref{Psi_def}).  
%---

Thus, looking at (\ref{density2}), 
we can see the following fact: 
When the chemical potential is increased 
with fixing all the other parameters, 
the particle density always grows. 
On the other hand, 
when the acceleration 
is decreased with fixing all the other parameters, 
the particle density always decreased.

%---
Thus, 
when decreasing the acceleration from higher one 
where there is no condensation, 
namely $\alpha=0$,
keeping the particle density to constant, 
the chemical potential should grow to keep the particle density constant. 
However, as can be seen from (\ref{bound_of_mu}), 
there is the upper bound for the value that the chemical potential can take, 
which is $m/a$. 
Thus, when the chemical potential reaches the upper bound, 
$\alpha$ should begin to grow.  
The finite value of $\alpha$ means 
the appearance of the condensation 
as written under (\ref{phi_def}). 
Thus, we can see that 
the critical value of the chemical potential is $m/a_c$. 

%%%%%%%%%%%%%%%%%%%%%%%%%%%%%%%%%%%%%%%%%%%%%%%%%%%%%%%
\subsection{Analysis in the relativistic situation}
\label{cueqw}
%%%%%%%%%%%%%%%%%%%%%%%%%%%%%%%%%%%%%%%%%%%%%%%%%%%%%%%

As mentioned under (\ref{summation-n}), 
our analysis is the one at the critical moment. 
Thus, $m/a_c$ is assigned to the value of the chemical potential 
with $\alpha=0$ and $\blue{\rho = a_c^{-1}}$. 
 
We put an assumption that $m/a_c$ is small. 
This means that the chemical potential 
in this study is also small. 
The situation that $m/a_c$ is small corresponds to the relativistic situation 
as mentioned in Sec.\ref{vnrver}. 

%-----------------------------------------
 
In what follows, 
we use two symbols, $\mu$ and $\mu'$. 
$\mu$ means the true chemical potential, 
where ``true'' means that it has been given 
at (\ref{Z_DpDp}) and the derivative with regard to $\mu$ can act on. 
On the other hand, $\mu'$ means just a value $m/a_c$. 
$\mu'$ is not the chemical potential and the derivative cannot act on.  

Then, it turns out that 
we can expand $\Phi$ and $\Psi$ with regard to $\mu$ and $\mu'$ as 
\begin{subequations}\label{Rel_Expansion1} 
\begin{align}
\label{Expanded_Phi}
\Phi =& \,\, \Phi_0+\Phi_2\,\mu^2+\cdots,\\
\label{Expanded_Psi}
\Psi =& \,\, \Psi_0\,{\mu'}^2+\Psi_2\,\mu^2+\cdots, 
\end{align} 
\end{subequations}
where 
\begin{subequations}
\begin{align} 
\Phi_0 \equiv& \,\, \frac{\red{1}}{\lambda} \, \coth (\pi \lambda), \\
%--- 
\Phi_2 \equiv& \, 
-\frac{1}{\red{2}\lambda^3} 
\Big( {\coth}(\pi \lambda)  
+ 
\frac{\pi \lambda}{\sinh^2(\pi \lambda)} 
\big( 1 - 2 \pi \lambda \, \coth(\pi \lambda) \big) 
\Big),\\
%---
\Psi_0 \equiv& \, - \! \Psi_2 \equiv a_c^2 \, K_{i \lambda}(k/a_c).
\end{align}
\end{subequations}
At this time, substituting (\ref{Expanded_Phi}) and (\ref{Expanded_Psi}) into (\ref{Gamma01}), 
$\Gamma$ at the critical moment in the relativistic situation is given as 
\begin{align}\label{Gamma02}
\Gamma = \, 
\frac{1}{2\blue{a_c}\pi^3} \int \! d \lambda \, \black{\lambda} \, \black{\sinh(\pi \lambda)} \int \! dk  \, \black{k} \left( \Phi_0\Psi_0\,\mu'{}^2+\Phi_0\Psi_2\,\mu^2 \right).
\end{align}

%%%%%%%%%%%%%%%%%%%%%%%%%%%%%%%%%%%%%%%%%%%%%%%%%%%%%%%
\section{The particle density and critical acceleration}
\label{urgei}
%%%%%%%%%%%%%%%%%%%%%%%%%%%%%%%%%%%%%%%%%%%%%%%%%%%%%%%

Now, we obtain the particle density $d = - \p \Gamma/\p \mu$ 
using (\ref{Gamma02}) as
\begin{align}\label{density01}
d =& \, - \frac{\mu}{\blue{a_c}\,\pi^3}\int_0^\infty d\lambda \lambda \sinh(\pi \lambda) \int \! dk k \, \Phi_0\Psi_2\nn
%=====
  =& \, \frac{\red{a_c\,\mu}}{\pi^3}\int_0^\infty d\lambda \, \cosh(\pi \lambda) \int \! dk k \, K_{i\lambda}^2(k/a_c)\nn
%=====
  =& \, \frac{\blue{a_c^3}\, \mu}{\red{2}\pi^2}\int_0^\infty d\lambda \, \lambda \, \coth \, (\pi \lambda),
\end{align}  
where we have used $\sinh(\pi \lambda) \, \Phi_0 = \red{\lambda^{-1}} \cosh(\pi \lambda)$, 
and performed the integrate with regard to $k$ as 
$\displaystyle \int dk k \, K_{i\lambda}^2(k/a) = \frac{ \pi \lambda}{2}\,a_c^2\,{\rm csch}(\pi \lambda)$.

Let us evaluate the $\lambda$-integral in (\ref{density01}): 
\begin{align}\label{ncpwc}
\int^\infty_0 d\lambda \lambda \coth (\pi \lambda). 
\end{align}
First, we can see that (\ref{ncpwc}) is diverged if it is evaluated as it is. 
Therefore, we consider to do some regularization toward (\ref{ncpwc})\footnote{
%=====================
%===== FOOTNOTE =====
%=====================
The divergence also appears in other works 
for the critical acceleration 
for the spontaneous symmetry breaking~\cite{Ohsaku:2004rv,Ebert:2006bh,Castorina:2012yg} 
and the $D=1+3$ Euclid space at finite temperature~\cite{Kapusta:2006pm}. 
In these, the divergences are ignored 
supposing that some regularizations could work, 
although it is not discussed explicitly. 
%=====================
}. 
Concretely, expressing $\coth (\pi \lambda)$ as
\begin{align}\label{nslvr}
\coth (\pi \lambda)= 1+\frac{2}{e^{2 \pi \lambda }-1}, 
\end{align}
we exclude ``$1$''. 
Then, once putting the upper limit of the integral as $\Lambda$, 
we get as
\begin{align}\label{nwsthe}
\int^{\Lambda}_0 d\lambda \lambda (\coth (\pi \lambda)-1)
=
-\frac{1}{12} 
-\Lambda ^2
+\frac{\Lambda \log (1-e^{2 \pi \Lambda })}{\pi }+\frac{\text{Li}_2(e^{2 \pi \Lambda })}{2 \pi ^2}.
\end{align}
Then, excluding ``$-\frac{1}{12}$'' in (\ref{nwsthe}), 
we take the $\Lambda$ to $\infty$. 
As a result, we can get as
\begin{align}\label{vdouyr}
\lim_{\Lambda \to \infty}
(-\Lambda ^2+\frac{\Lambda \log (1-e^{2 \pi \Lambda })}{\pi }+\frac{\text{Li}_2(e^{2 \pi \Lambda })}{2 \pi ^2})=\frac{1}{6}. 
\end{align}

As a result, 
by putting $\mu = m/a_c$ as the critical moment, 
we can obtain the particle density as
$d = 
{\blue{a_c^2} \, m}/{12\pi^2}$.
From this, 
we can obtain the critical acceleration as
\begin{align}\label{result}
a_c = 2 \, \pi \,\sqrt{{3d}/{m}}.
\end{align}
Thus, the critical Unruh temperature is obtained as $\sqrt{{3d}/{m}}$, 
which agrees with the usual result in the 4D Euclid space at finite temperature.

%%%%%%%%%%%%%%%%%%%%%%%%%%%%%%%%%%%%%%%%%%%%%%%%%%%%%%%
\section{Summary}
%%%%%%%%%%%%%%%%%%%%%%%%%%%%%%%%%%%%%%%%%%%%%%%%%%%%%%%

In this study, based on the thermal excitation by the Unruh effect, 
we have analyzed the critical acceleration at which the BEC is formed as in (\ref{result}).   
%-----
The system we have considered is an accelerating ideal gas composed of a complex scalar particle 
in the Minkowski spacetime at zero temperature. 
For this purpose, 
we have considered a free complex 
scalar field at finite density given in (\ref{action}) 
in the Rindler coordinate system given in (\ref{Rindler_metric}). 

In our analysis, 
we have picked up the leading term 
supposing that 
the value of the critical chemical potential 
is small as in (\ref{Rel_Expansion1}). 
This means to assume the relativistic situation, 
$k_B T/2 \gg mc^2$ (which corresponds to the situation 
that the energy of particle's thermal motion is very larger than its static energy). 
The analysis in the non-relativistic situation, $k_B T/2 \ll mc^2$,   
will be tried in the future work. 

In this study, there has been some problem regarding the divergence and its regularization as in Sec.\ref{urgei}. 
The analysis in this work has not gone into its details. 
%Also, there is the problem, 
%which is whether thermal phenomena 
%arisen by the Unruh effect in the accelerated frame 
%can be observed or not 
%in the inertial frame. 
%We have performed this work 
%supposing it occurs. 
%However there is a study to argue that 
%the Unruh effect does not contribute to the restoration of 
%the spontaneous symmetry breaking~\cite{Hill:1985wi}. 
We will give our consideration for these problems in the future work. 

%%%%%%%%%%%%%%%%%%%%%%%%%%%%%%%%%%%%%%%%%%%%%%%%%%%%%%%

\end{document}